\documentclass[preprint,authoryear,12pt]{elsarticle}

\usepackage{epsfig}

\usepackage{amssymb}

\usepackage[ps2pdf,%
a4paper=true,%
breaklinks=true,%
colorlinks=true,%
pdfauthor={First Author et al.},%
pdftitle={Template for manuscripts in Advances in Space Research}%
]{hyperref}

\journal{Advances in Space Research}

\begin{document}

\begin{frontmatter}



\title{Modeling D-Region Ionospheric Response of the Great American TSE of August 21, 2017 from VLF signal perturbation}


\author[label2,label1]{Sandip K. Chakrabarti}
\ead{sandipchakrabarti9@gmail.com}
\author[label1]{Sudipta Sasmal\corref{cor}\fnref{footnote1}}
\ead{meet2ss25@gmail.com}
\cortext[cor]{Corresponding author}
\fntext[footnote1]{Tel: 91-33-2436 6003; Ext. 21}
\author[label1]{Suman Chakraborty}
\ead{suman.chakrabarty37@gmail.com}
\author[label1,label3]{Tamal Basak}
\ead{tamalbasak@gmail.com}
\author[label4]{Robert L. Tucker}
\ead{rltucker@aol.com}
\address[label1]{Indian Centre for Space Physics, 43 Chalantika, Garia Station Road, Kolkata - 700084}
\address[label2]{S. N. Bose National Centre for Basic Sciences, JD Block, Sector-III, Salt Lake, Kolkata - 700098}
\address[label3]{Department of Physics, AIAS, Amity University, Major Arterial Road, Action Area II, Rajarhat, New Town, Kolkata - 700135, India}
\address[label4]{7911 E 26 Place, Tulsa, OK, 74129, USA}


\begin{abstract}
\noindent{Solar eclipse is an unique opportunity to study the lower ionospheric variabilities under a controlled perturbation when the solar 
ultraviolet and X-ray are temporally occulted by the lunar disk. Sub-ionospheric Very Low Frequency (VLF) radio signal displays the ionospheric 
response of solar eclipse by modulating its amplitude and phase. During the Total Solar Eclipse (TSE) on August 21, 2017 in North America, data was 
recorded by a number of receivers as presented in public archive. Out of these, two receiving stations YADA in McBaine and K5TD in Tulsa could 
procure a reasonable quality of noise free data where the signal amplitude was clearly modulated due to the 
eclipse. During the lunar occultation, a C3.0 solar flare occurred and the signal received from Tulsa manifested the effect of sudden ionization due 
to the flare. The VLF amplitude in Tulsa shows the effect which is generally understood by superimposing effects of both the solar eclipse and flare. 
However, the signal by YADA did not perturb by the solar flare, as the flaring region was totally behind the lunar disk for the entire period. We 
numerically reproduced the observed signal amplitude variation at both the receiving locations by using Wait's two component D-region ionospheric 
model and the well-known Long Wavelength Propagation Capability (LWPC) code. The perturbed electron density for both the cases is computed which 
matches satisfactorily with the true ionospheric conditions.}
\end{abstract}

\begin{keyword}
Solar eclipse; Solar flare; Lower ionospheric anomalies; VLF; LWPC.
\end{keyword}

\end{frontmatter}

\parindent=0.5 cm

\section{Introduction}

\noindent{Earth's ionosphere is a tenuous electrified layer of the atmosphere that responds to slight changes coming from both below and above. Such 
changes induce perturbations in the ionosphere that in turn interfere with radio signals propagating within the Earth-ionosphere waveguide (EIWG) and 
gets manifested as anomalies in the received signal amplitude and phase. Solar eclipse provides an unique opportunity to study such mechanisms behind 
ionospheric variations as its time of occurrence is pre-determined and thus it gives chance to equip ourselves accordingly [Crary and Scheneible 1965; 
Mitra 1974; Sen Gupta et al. 1980; Lynn 1981; Pant and Mehra, 1985; Thomson 1993; Clilverd et al. 2001; Kozlov et al. 2007; Karimov et al. 2008; 
Chernogor 2010; Chakrabarti et al. 2010, Maji et al., 2012; Pal et al., 2015; Chakraborty et al., 2016]. During a solar eclipse, the Moon comes in 
between the Sun and the Earth for a brief period, resulting a sudden loss of solar radiation over a specific region on the Earth. Without the 
ionizing radiation, the ionosphere tries to relax from the daytime to the nighttime condition and then goes back to normal situation after the eclipse 
is over. It thus replicates a day-night condition in a small region on the Earth in a quick succession. As a result of this, the ionospheric 
reflection height increases at the onset of the eclipse and decreases in the declining phase smoothly affecting the received signal amplitude and 
phase. Just as a solar eclipse causes reduction in ionizing radiation for a short period, a solar flare exhibits exactly the opposite nature where 
enhanced X-ray flux causes a sudden and rapid increase of ionization in the lower regions of the ionosphere [Tsurutani et al., 2009]. Such injection 
of energized particles into the ionosphere disrupts the normal balance of ion formation and recombination and disturbs the propagation of radio waves 
through it [Thome and Wagner, 1971; Mitra, 1974; Donnelly, 1976; Tsurutani et al., 2009; Palit et al., 2013; Basak et al., 2013].}\\

\noindent{A total solar eclipse occurred in North America on August 21, 2017 with a totality band extending from the Pacific coast to the Atlantic
coast. The lunar shadow began to cover land as a partial eclipse at 15:46:48 UTC. The total eclipse began at 16:48:32 UTC, being maximum at
18:26:40 UTC and ending at 20:01:35 UTC. The partial eclipse ended at 21:04:19 UTC. During this period, two solar flares of class C3.0 and C1.5
occurred at 17:57:00 UTC and 20:12:00 UTC respectively. The excess solar irradiance from these two flares were fully or partly occulted by the lunar 
disc and it offered us an excellent opportunity to probe the ionosphere under the influence of two opposing factors. Previously, similar observations 
were reported in Maji et al. (2012) and Pal et al. (2012) at the time of the annular solar eclipse (maximum obscuration $75\%$) that took place on 
January 22, 2010. During that eclipse, the Sun was exhibiting a flaring activity that reached its peak (C-type) immediately after the eclipse maximum. The 
receiving location was Khukurdaha (Lat: 22$^\circ$27$'$N, Long: 87$^\circ$45$'$E) and the signal monitored was from NWC transmitter transmitting at 
19.8 kHz. In Maji et al., (2012), the VLF signal amplitude during the period of the eclipse was reconstructed using analogies with previous 
observations and the best fit parameters were determined to reproduce the signal variation both in the presence and absence of the occulted flare. In 
Pal et al., (2012), the time variation of electron density profile under the combined influence of the eclipse and the occulted flare was 
theoretically calculated and compared with a normal solar flare.}\\

\noindent{In this paper, contrary to the previous work where such event was first reported during an annular eclipse, this observation is made during a total solar 
eclipse and the receiving locations were far away from the Indian landmass. Data from two receiving locations in North America, namely, YADA (McBaine)
and K5TD (Tulsa) are studied of which McBaine was within the totality belt with $100\%$ obscuration and so could not see the 
flares while Tulsa was close to totality ($88.61\%$) and experienced the flares. The theoretical approach of studying the ionospheric modulations 
under the combined effects of the eclipse and the flares starts with the calculation of the obscuration function. For Tulsa, this function was further 
modified to include the effects of the flares. Using this information, the Wait's exponential ionospheric parameters, $h'$ (ionospheric reflection height) and $\beta$ 
(steepness parameter) are then calculated and are used in the Long Wavelength Propagation Capability (LWPC) code to obtain the desired signal 
amplitude variation. The electron density profiles are also calculated for these two places, both including and excluding the effects of the 
flares.}\\

\noindent{The plan of the paper is as follows: In Section 2, we present the observational setup and data used in the study; in Section 3, we the methodology used 
in the study; in Section 4, the results obtained from numerical simulation and finally in Section 5, we will make concluding remarks.}

\section{Observation and Analysis}

\noindent{In this manuscript we use VLF data from one antenna-receiver system from SuperSID network of Stanford University Solar Center (http://solar-center.stanford.edu/SID) and one inverted L antenna with an HF amateur radio receiver (referred to as `Tulsa receiver'). 
During the Total Solar Eclipse (TSE) on August 21, 2017, an active network of SuperSID users and others 
monitored the ionospheric response of TSE in the North American region from several monitoring stations which
were chosen in such a way that the VLF signal can be received from above, below and on the totality belt. We use publicly available 
data from the SuperSID website (http://sid.stanford.edu/database-browser/). We choose one receiving location with Tulsa receiver
(Lat. 36.13$^\circ$N, Long. 95.89$^\circ$W) which is received by one of us (RLT) and a superSID receiver at McBaine (Lat. 38.90$^\circ$N, 
Long. 92.39 $^\circ$W) for our analysis. We use the VLF signal transmitted from NML transmitter from La Moure, North Dakota (Lat. 46.37$^\circ$N, Long. 
98.33$^\circ$W) of frequency 25.2 kHz. The great circle distance (GCP) between NML-K5TD (Tulsa) and NML-YADA (McBaine) are 1157 km and 962 km 
respectively. McBaine was sitting on the totality belt whereas Tulsa was below the belt of totality.}\\

\noindent{The SuperSID system contains a typical loop antenna connected with a pre-amplifier system. The output of the pre-amplifier is connected 
with a sound card or data acquisition card attached with computer for analog to digital signal conversion. The acquisition, processing and storing of 
the data are done by a computer programme provided with the SuperSID instrument. The Tulsa receiver has a HF amateur radio receiver with an inverted L
antenna with the vertical length of 35ft and a horizontal (E-W) length of 30ft. In Fig. 1(a-c) the location of NML, Tulsa and McBaine are 
marked. The red curve shows the central line of the totality belt which is the trajectory of all the locations from where the TSE (100\% obscuration) 
was experienced. The colored patch is the lunar shadow as projected on Earth through the center of which the central line passes. The Fig. 1a-c 
are for starting of the TSE, maximum eclipse and ending of TSE. The color-bar represents the luminance of incoming solar flux which is just 
inverse of the solar obscuration function. It is clear that at the center of the totality line, the value of the obscuration function is 1 which 
means that one sees the Sun being totally occulted by the Moon from this region. The calculation of the obscuration function is given in details in 
the next Section.}\\

\begin{figure}[h]
\centering{
\includegraphics[width=6.5cm]{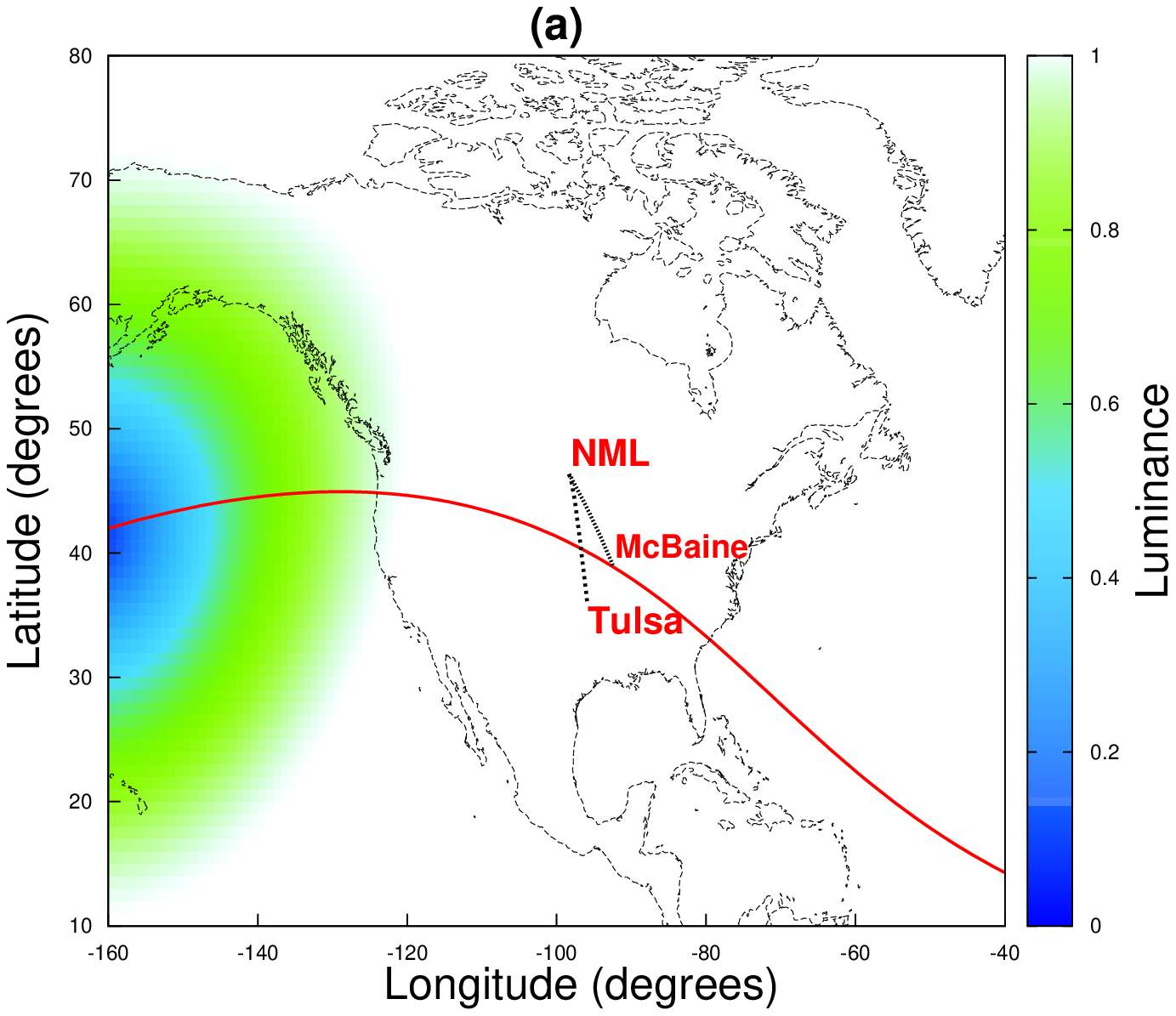}
\includegraphics[width=6.5cm]{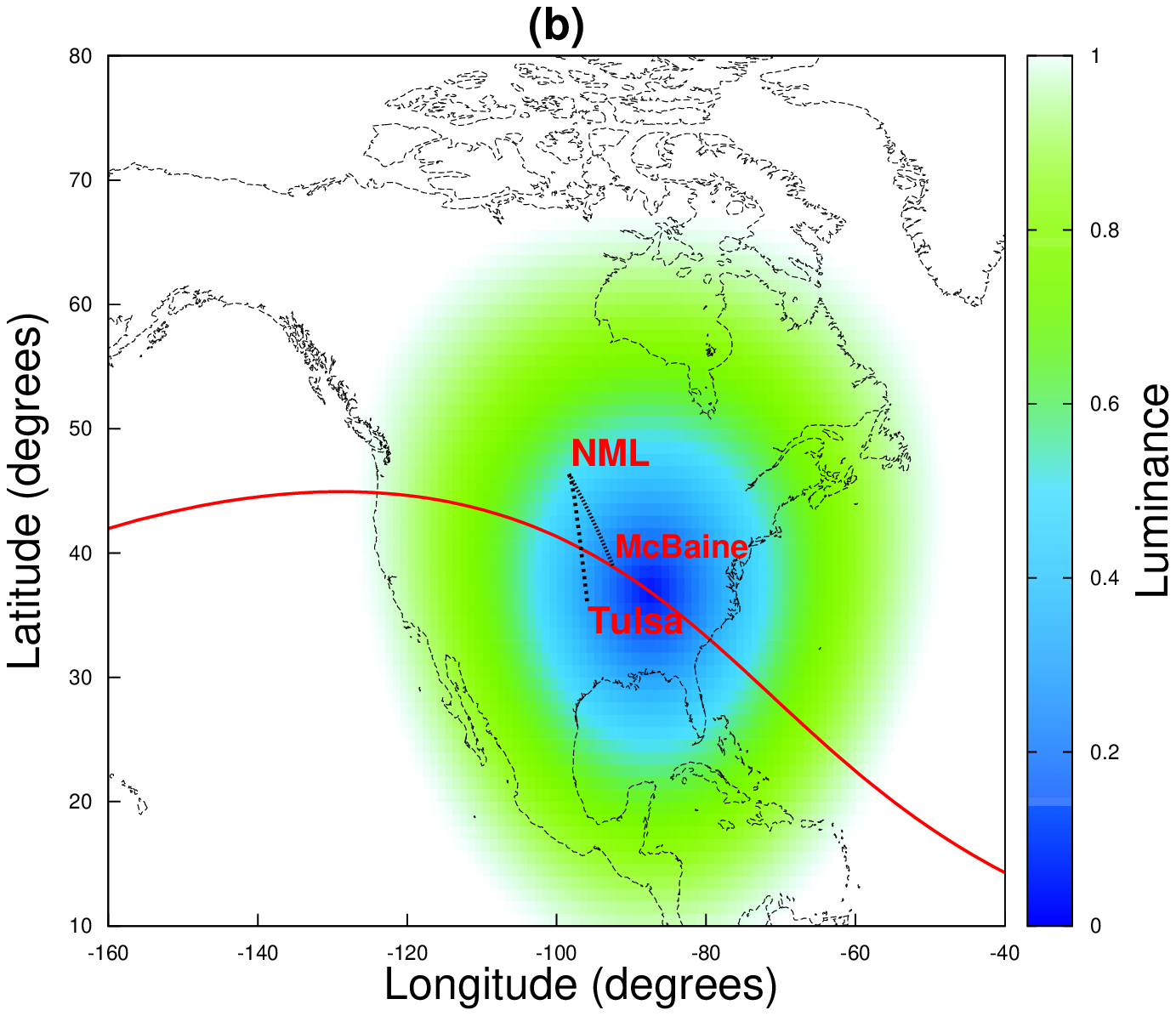}
\includegraphics[width=6.5cm]{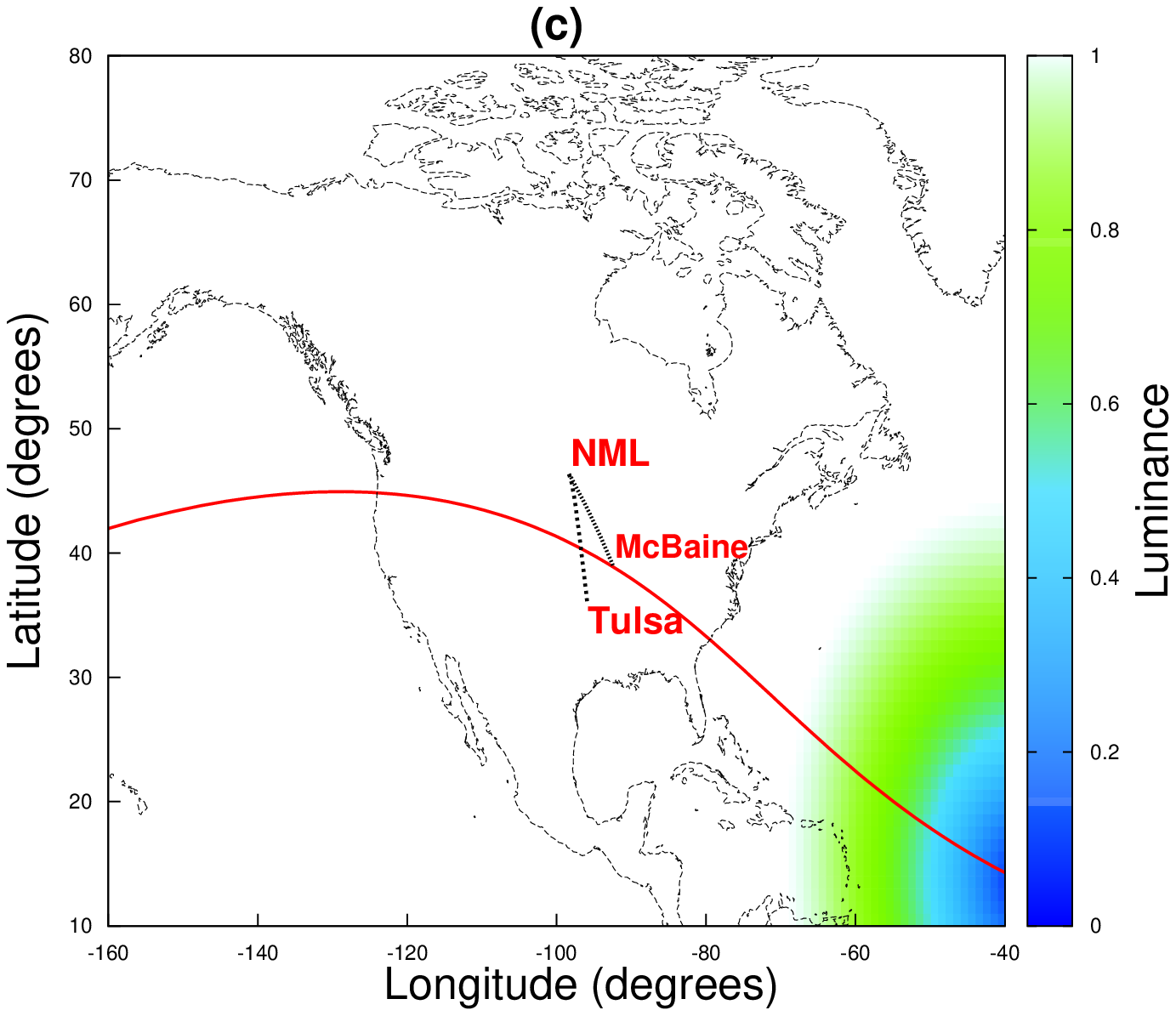}}
\caption{The location (La Moure) of the transmitter NML, the receiving locations McBaine and Tulsa 
and the two propagation paths marked with black lines. The path 
of the totality is marked by the red curves. The Figures (a), (b) and (c) represents the starting, maximum and ending of the total solar eclipse 
respectively. The color-bar represents the luminance where 0 denotes full occultation and 1 denote full luminance.}
\end{figure}

\clearpage

\noindent The detailed information of the eclipse at the two receiving locations are given in Table I.

\begin{table}[h]
\caption{Information of Eclipse}
\centering
\small
\begin{tabular}{|c|c|c|c|c|}
\hline
Receiving Location & Starting & Maximum & Ending & Obscuration (\%)\\
\hline
McBaine & 16:45 & 18:13 & 19:40 & 100\\
\hline
Tulsa & 16:39 & 18:08 & 19:37 & 88.61\\
\hline
\end{tabular}
\end{table}

\noindent{Maji et al. (2012) reported the first ever observation of effects of solar flare during an eclipse. The present paper is unique in the sense 
that during the TSE two solar flares, namely C3.0 and C1.5 occurred. One place (Tulsa) could see a partial blocking of one flare. This was possible 
since the maximum obscuration of the solar disc was 88.6\%.  During the other flare the eclipse was already over. The other place (McBaine) could not see 
the effects of both the flares.  The lunar disc was totally blocking the Sun during the C3.0 and the eclipse was over during C1.5. During the eclipse, 
a nighttime condition happens when the D-region electron density decreases. During a flare, on the other hand, it is exactly the opposite. Figure 2c 
shows the GOES15 1$\AA$ - 8 $\AA$ X-ray light curve of the C3.0 class solar flare that occurred during the solar eclipse.}\\ 

\noindent{Figure 2(a,b,c) shows the ionospheric response of VLF signal amplitude variation as a function of time in hours for NML-YADA (Top) and 
NML-K5TD (Middle) and the GOES X-ray lightcurve of the C3.0 solar flare. Originally the recorded signals had the unit of millivolts as calibrated in the 
SuperSID instrument. To make use of the data and for theoretical reproduction we normalized the signal amplitude by using the famous Long Wavelength 
Propagation Capability (LWPC) programme to get the signal amplitude in the dB format. We choose the time duration of 4 hours (16:00:00 UT to 20:00:00) for signal 
representation. The vertical dashed lines are the starting, maximum and ending of the eclipse with respect to the location of the receiver.}\\

\begin{figure}[h]
\centering{
\includegraphics[width=12.0cm]{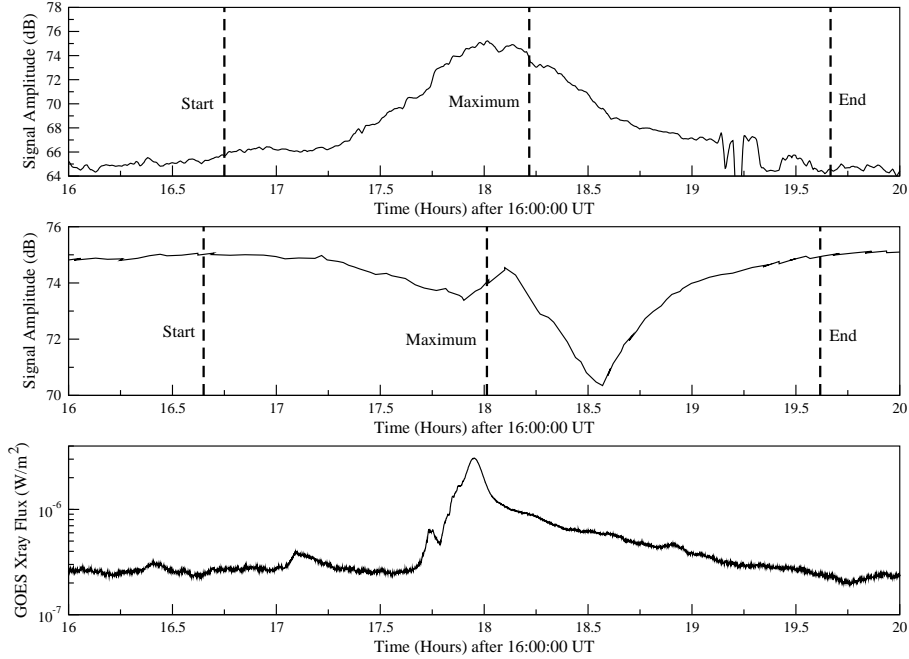}}
\caption{The VLF signal amplitude variation as a function of time in hours in UT for (a) NML-YADA and (b) NML-K5TD path during the total solar eclipse 
on August 21, 2017. The three vertical dashed thick lines represent the starting, maximum and ending of the total solar eclipse. In (c) we
show the GOES15 $1\AA$ - 8 $\AA$ X-ray light curve of C3.0 class flare in the same time slot.}
\end{figure}

\noindent{The VLF response in Tulsa and McBaine due to the eclipse are completely opposite in nature. It is clear from Fig. 2 that 
for McBaine, the signal amplitude increases due to the eclipse. On the contrary, for Tulsa, originally the signal amplitude starts to decrease. After 
that there is a sudden rise and fall and then again it starts to increase. This sudden enhancement of the signal amplitude and a decay is due to the 
solar flare. If as in McBaine, the obscuration were 100\% then there would have been a completely opposite nature of signal amplitude for Tulsa in 
comparison with McBaine as we could show from our model described below.}

\section{Methodology and Modeling}

\noindent{The NML-K5TD and NML-YADA are two short VLF propagation paths over which the totality belt of shadow of TSE-2017 has gone through 
(Fig. 1). Hence, the signatures of eclipse induced lower ionospheric perturbations are imprinted on the VLF signal amplitudes. Though those paths are 
geographically not at all far away (Fig. 1) and the difference in geographic bearing angle between them is approximately $21^\circ$ (Fig. 1). So, due to 
the characteristic anisotropy of the transmitted signal from a NML transmitter and the asymmetry among the mutual interferences of signal propagation 
modes over those paths, we observe totally opposite types of perturbations in recorded VLF signal amplitudes. The signal is enhanced 
($+ve$ type) in NML-YADA and and attenuated ($-ve$ type) in NML-K5TD path (Fig. 2). Most interestingly the C3.0-class solar flare which 
occurred entirely within the eclipse period and was detected only by VLF-observation at NML-K5TD and not by NML-YADA path. This has happened because 
the flaring region of the Sun was occulted partially or fully by the lunar disc as viewed from NML-YADA path during the flare occurrence time. As a 
result, the perturbed signal of NML-K5TD path actually carries the composite effect of solar eclipse and the solar flare. In our analysis, we assumed it to 
be a linear superposition of these two effects so that separate contributions from each of those effects can be subtracted out. Pal et al. (2012) and 
Maji et al. (2012) have applied this linear superposition assumption and obtained satisfactory results for VLF-observation analysis during annular 
solar eclipse of 2010.}\\

\noindent{To model the lower ionospheric profile over the signal propagation path, we use the Long Wave Propagation Capability (LWPC) code (Ferguson, 
1998). The default propagation model of LWPC for numerically simulating the propagating VLF signal amplitude within the earth and ionosphere 
cavity is Long Wave Propagation Model (LWPM). LWPM assumes a simplified model of D-region ionosphere having exponentially increasing conductivity 
profile with height (Ferguson, 1998; Pal et al., 2010; Sasmal et al., 2017). Apart from LWPM, LWPC uses a number of specialized models, namely, 
HOMOGENEOUS, CHI, RANGE, and GRID. The objectives of the models are to introduce several perturbation effects into the ionosphere under different 
circumstances. The specific inputs can be fed to the models using the inbuilt substrings of LWPC, namely TABLE and EXPONENTIAL. In our simulation 
here, we use the RANGE model and the EXPONENTIAL substring.  The parameters of this model are called effective signal reflection height (h$^\prime$) 
and sharpness factor (the log-linear slope of D-region electron density profile) $\beta$. The global conductivity map ($\sigma$) and geomagnetic 
field values are taken from inbuilt database of LWPC. The altitude profile of lower ionospheric electron density and electron-neutral collision 
frequency is obtained from Wait's 2-component ionospheric model using these above mentioned parameters h$^\prime$ and $\beta$ (Wait and Spies, 
1964).}\\

\noindent{Now, we know that the VLF signal raw voltage data are recorded in an arbitrary unit of the receiver systems for both of those stations. 
Since, we use LWPC code for simulation and it calculates the signal amplitude in $dB$, we have to normalize the raw data to logarithmic decibel 
scale. This normalization of signal would not affect the results because in this work, we are particularly interested in understanding and 
simulating the relative changes in signal amplitude due to lower ionospheric perturbations by solar eclipse. Before doing this normalization, we have 
chosen the Wait's parameters for unperturbed D-region ionospheric profile for the two paths to be $h^\prime=70$ km and $\beta=0.3$ km$^{-1}$ 
according to the prescription of Thomson (1993). As we are simulating the VLF signal profile for comparatively short propagation paths, we 
purposely neglect solar zenith angle ($\chi$) variation over the paths and we assume a uniform solar radiation induced ionospheric ionization 
profiles over those paths. In Pal et al., (2012), similar exercise was done for VTX-Kolkata (1946 km), VTX-Malda (2151 km), VTX-Raigunj (2207 km) etc. 
paths during TSE-2009. We run the RANGE-EXPONENTIAL model of LWPC using the parameter values corresponding to the unperturbed D-region conditions 
for both NML-Tulsa and NML-YADA paths and the obtained VLF signal amplitudes at the receiving points are 75 dB and 65.1 dB respectively. For 
normalization we use the following formula:}

\begin{equation}
A(t) = 10\, log\frac{I(t)}{I_0},
\end{equation}

\noindent{where, $I(t)$ is the VLF amplitude recorded at receiving stations in arbitrary units, $A(t)$ is the normalized amplitude of VLF in 
$dB$ (used in LWPC), and $I_0$ is the normalization factor. Since the eclipse over those paths started at around 16:40:00 UT, we choose a time 
16:00:00 UT which is well before its commencement and hence, it can be justifiably assumed that the D-region was more or less in unperturbed 
condition. We take the raw data amplitude ($I(t$=16:00:00 UT)) at that time and normalize them to LWPC simulated values 75 dB and 65.1 dB 
($A(t$=16:00:00 UT)) respectively using Eqn. 1. Thus, we obtained the normalization factors ($I_0$) each for those two stations. Using the respective 
$I_0$ values in Eqn.1, we normalize the entire raw data set in $dB$ units (Fig. 2). We get the maximum signal changes (i.e. the differential 
amplitude, $\Delta A_{max}$) of +10.1 $dB$ for NML-YADA and $-4.5$ dB for NML-K5TD paths.}\\

\noindent{We first perform the LWPC simulation at the peak region of perturbation and obtain the change in effective reflection heights 
($\Delta h^\prime_{max}$) over the paths to +10.0 km and +4.0 km respectively. These correspond to the differential amplitudes ($\Delta A_{max}$) 
of +10.1 dB (NML-YADA) and -4.5 dB (NML-K5TD) respectively. The complete nighttime situation near the D-region ionosphere is characterized by 
h$^\prime$=87 km (Ferguson, 1998). NML-YADA path was across 100\% obscuration zone. However, the highest value of perturbed h$^\prime$ is (70+10.0) = 
80 km, i.e. 59\% of the total nighttime situation has occurred on the NML-YADA path during totality. On the other hand, NML-K5TD path experienced a 
maximum of 88.6\% obscuration and the height est perturbed h$^\prime$ is ($70+4$) = $74$ km, i.e., only 23.5\% of the total nighttime situation 
occurred in spite of 88.6\% obscuration by lunar disc. This observation can be explained by emission properties of the X-ray and other ionizing 
radiations from the solar corona. This solar corona may extend up to a distance of a few solar radii from the solar photosphere. Indeed, TSE-2017 had 
a spectacular show of extended corona all around the solar disc. Thus even during the totality, radiation coming solely from the corona can and do 
ionize the D-region significantly. To incorporate the effect of C3.0-class flare, we use the soft X-ray (1-8$\AA$) light curve of the same from 
GOES-15 satellite data (Fig. 2). As a first order approximation, we assume a linear variation of Wait's parameters with X-ray energy (Basak and 
Chakrabarti, 2013; Pal and Chakrabarti, 2010).  We then run the LWPC-scheme and scale the values of $h^\prime$ and $\beta$ in order the appropriately 
simulate the observed VLF-amplitude part due to flare effect. Using this mechanism, we simulate the Wait's parameter values and obtain the entire 
VLF signal amplitude profile $A(t)$. In the next step, we put the h$^\prime$ and $\beta$ values to the famous Wait's formula (Wait and Spies, 1964) to 
obtain the temporal variation of altitude profile of D-region electron density (N$_e$(h,t)) during TSE-2017 for both NML-YADA and NML-K5TD paths. To 
model the solar eclipse induced perturbation of D-region ionosphere, first we have to estimate the degree of solar obscuration over the signal 
propagation path. The degree of solar obscuration is defined as the ratio between the Sun's projected area blocked by the Moon to the Sun's total 
projected area as seen from the Earth [Chakraborty et al., 2016]. In Fig. 3, we draw the Sun-Moon 
configuration during an eclipse. If $r$ and $R$ are the radii of the Sun and the Moon respectively as seen from the central line of the eclipse shadow,
$\theta$ and $\phi$ are the angles subtended by the points of intersection of the two perimeters at the respective centers, $D$ is the distance 
between the two centers and $d$ is the distance of the overlapping region, then the degree of obscuration $(p)$ will be given by,}

\begin{equation}
p = \frac{r^2(\theta-\cos\theta\sin\theta) + R^2(\phi-\cos\phi\sin\phi)}{\pi r^2}.
\end{equation}

\noindent{Where}, 

$$\cos\theta = \frac{1}{2}\left(\frac{r^2-R^2+(R+r-d)^2}{r(R+r-d)}\right),$$
\noindent{and},
$$\cos\phi = \frac{1}{2}\left(\frac{R^2-r^2+(R+r-d)^2}{R(R+r-d)}\right).$$

\begin{figure}[h]
\centering{
\includegraphics[width=12.0cm]{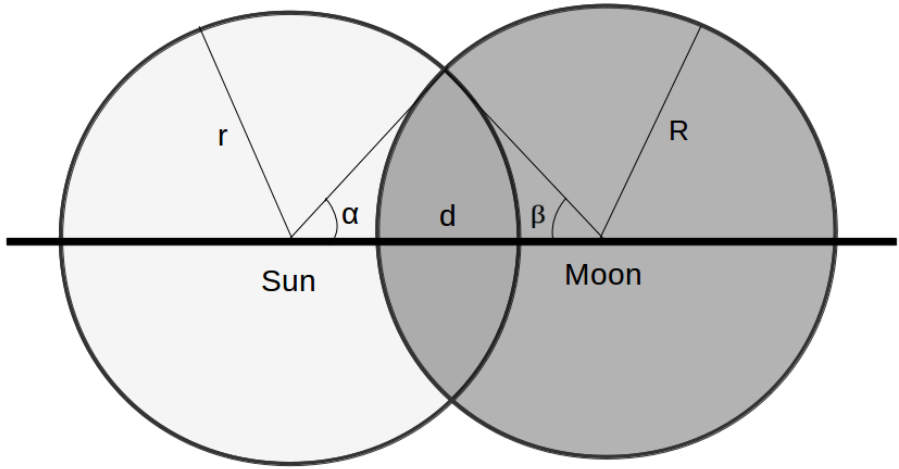}}
\caption{Geometrical representation of the Sun and the Moon during a solar eclipse [Mollmann and Vollmer, 2006].}
\end{figure}

\noindent{The angular diameter of the Sun and the Moon appears to be almost equal when seen from Earth, specially when seen from within the totality
belt and hence we can take r and R to be equal. This approximation makes the cosine functions to be equal as well

$$\cos\theta = \cos\phi = \frac{1}{2}\frac{D}{r}$$,

\noindent{and Eq. 2 then modifies to,}

\begin{equation}
p = \frac{2}{\pi}(\theta - \cos\theta\sin\theta).
\end{equation}

\noindent{Luminance (L) at any point on Earth under shadow region is exactly the opposite of obscuration function (p) and is defined as}

$$\frac{L}{L_{max}} = 1-p,$$

\noindent{where $L_{max}$ is the maximum luminosity of the Sun and taking it to be 1. We can rewrite luminosity as}

$$L = 1-p.$$

\noindent{Clilverd et al., (2001) and Pal et al., (2012) assumed a linear variation of Wait's parameters (both h$^\prime$ and $\beta$) with the 
calculated obscuration function. For signal analysis over longer propagation paths this approximation may not be acceptable in general (Lynn 1981, 
Patel et al., 1986). In this paper, as we are dealing with short paths ($\sim$1000 km) (Fig. 1) the approximation could be acceptable. To separate 
the effect of the solar flare, we supply $\Delta h^\prime$ and $\Delta \beta$ values in linear proportion to the calculated fractional obscuration 
values to LWPC. Using this method, we calculate the resulting VLF amplitude perturbation due to TSE-2017 ($\Delta A(t)$) alone over the NML-K5TD path. 
Again we use the the set of Wait's parameters for eclipse effects only to obtain the temporal profile of modeled electron density (N$_e$(h, t)).

\section{Results and Interpretations}

\noindent{In this Section, we present the results of our numerical modeling. Figure 4 shows a comparative study of observed and simulated signal 
amplitude as a function of time in hours for the path NML-YADA. The topmost panel (a) shows the observed signal variation, the middle panel (b) shows 
the simulated signal amplitude and the bottom panel (c) shows the degree of obscuration at the receiving location. In all the panels, along X-axis, 
we plot the time (UT) in hours. In panels (b) and (c), along Y-axis, we plot the signal amplitude in dB. In panel (c), we plot the degree of 
obscuration which varies from $0$ (no obscuration) to $1$ (full obscuration). From Fig. 4a, we find that the signal amplitude shows a general trend of 
positive deviation from normal unperturbed values during the eclipse period and the results obtained from simulation as shown in Fig. 4b exhibits a 
similar trend. Also, the time of maximum deviation (18:16:00 UTC) as obtained from simulation matches observation (18:13:00 UT) to a great extent.}\\

\begin{figure}[h]
\vskip 0.8cm
\centering{
\includegraphics[width=12.0cm]{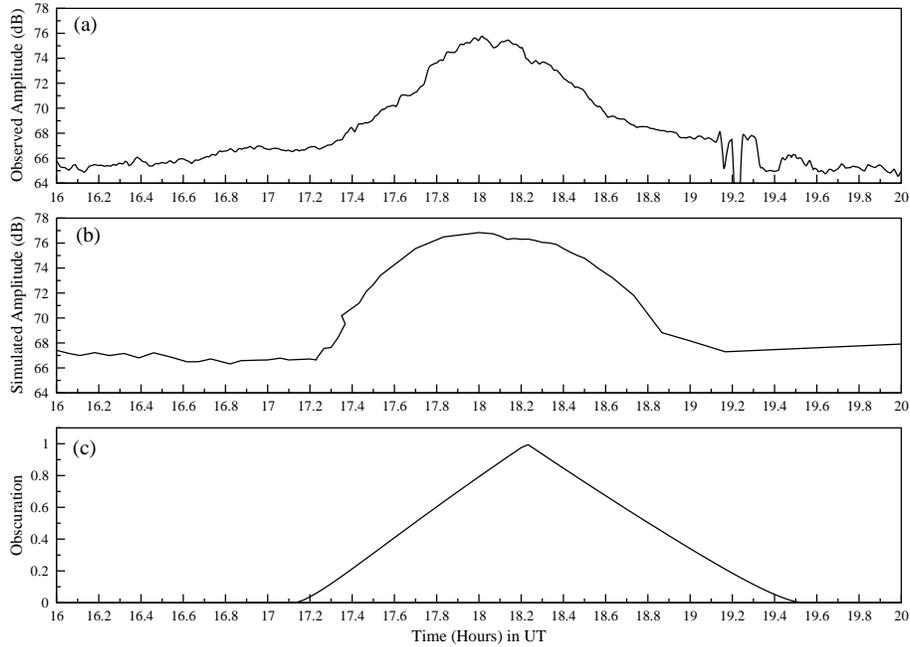}}
\caption{Comparison of the observed and modeled signal amplitude as a function of time in hours for NML-YADA path. The top column (a) shows the 
observed signal, middle (b) shows the modeled amplitude and the bottom (c) shows the obscuration function. The observed and modeled signal 
amplitude matches satisfactorily.}
\end{figure}

\noindent{Figure 5 shows a similar comparison of the signal amplitude variation during eclipse time for the path NML-K5TD. The signal amplitude for 
this path exhibits an important additional feature due to the occurrence of a solar flare during the eclipse and produced a combined effect on the 
received signal amplitude. From Fig. 5a, we see that the signal amplitude begins to show a negative deviation with the start 
of the eclipse but makes a quick reversal in the trend with the onset of the solar flare and then again begins to decrease. Exactly similar trend of 
signal modulation is obtained from simulation as can be seen from Fig. 5b. The time of maximum deviation (18:36:00 UTC) is also found to match 
observation (18:34:00 UT) quite satisfactorily. Here, we clearly see that although the time of maximum obscuration at Tulsa was 18:06:00 UT, the 
maximum deviation registered in the received VLF signal amplitude is at 18:34:00 UT. This delay of $\sim 28$ minutes is a clear manifestation of the 
effects of the flare. Previously, similar delay of 7 minutes was reported in Maji et al., [2012] but in our case, we obtained the effects to be more
strong which is due to the strength of the affecting flare (C3.0).}\\

\begin{figure}[h]
\vskip 0.8cm
\centering{
\includegraphics[width=12.0cm]{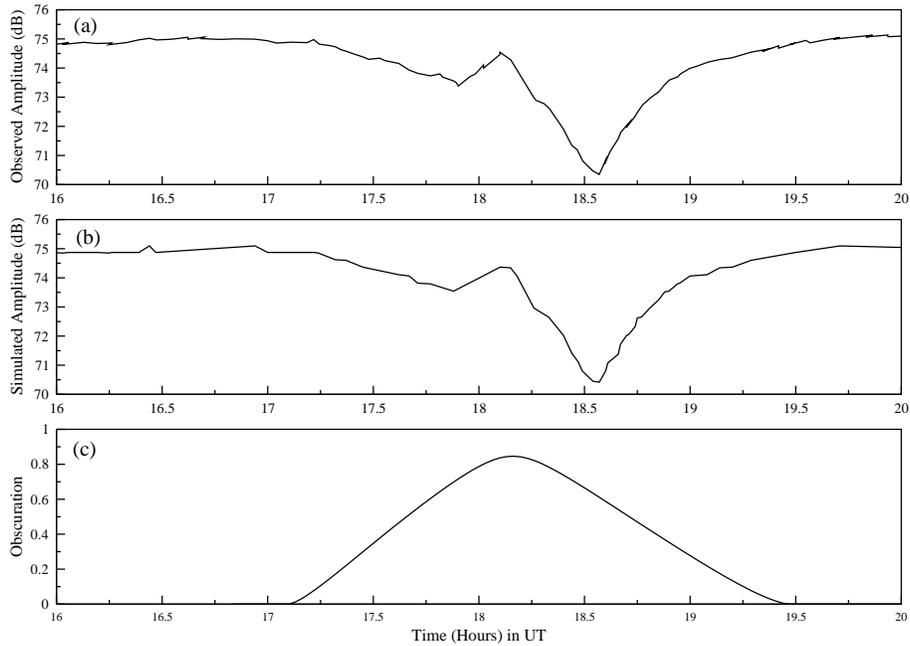}}
\caption{Similar comparative study as like Figure 4 for NML-K5TD path.}
\end{figure}

\noindent{In Figure 6, we show the simulated signal amplitude for the path NML-K5TD. Along X-axis, we plot the time (UT) in hours and along Y-axis
we plot simulated signal amplitude in dB. The blue curve corresponds to the situation if there were no solar flare and the signal amplitude were
modulated solely due to the solar eclipse. The red curve corresponds to the situation where both the solar eclipse and the solar flare modulate
the signal amplitude. The two vertical dotted lines denote the time of minimum signal amplitude as obtained from simulation. The ionosphere is
maintained by a series of complex photochemical reactions governing primarily the production and recombination of different species of ions and 
electrons. A solar eclipse ceases the ionizing radiation for a brief moment of time thereby accelerating the recombination processes leading to 
the loss of ions and electrons. A solar flare plays an exactly opposite role and speeds up the ionization process. As in the present study, the 
flare began right after the starting of the eclipse and the ionosphere exhibits a sluggishness in its response to the recombination processes. The 
signal amplitude displayed an exciting feature in its modulation during that period. From simulation also, we have been able to reconstruct the
same trend. From Fig. 6, we see that if there were no flare and the eclipse alone had influenced the received signal amplitude, then the minimum 
would have been achieved at 18:24 UT. In the actual situation where both the flare and the eclipse had influenced the signal amplitude, the
minimum is achieved at 18:36 UT. This clearly implies a delay of obtaining minimum in the signal amplitude by 12 minutes.}

\begin{figure}[h]
\vskip 0.8cm
\centering{
\includegraphics[width=10.0cm]{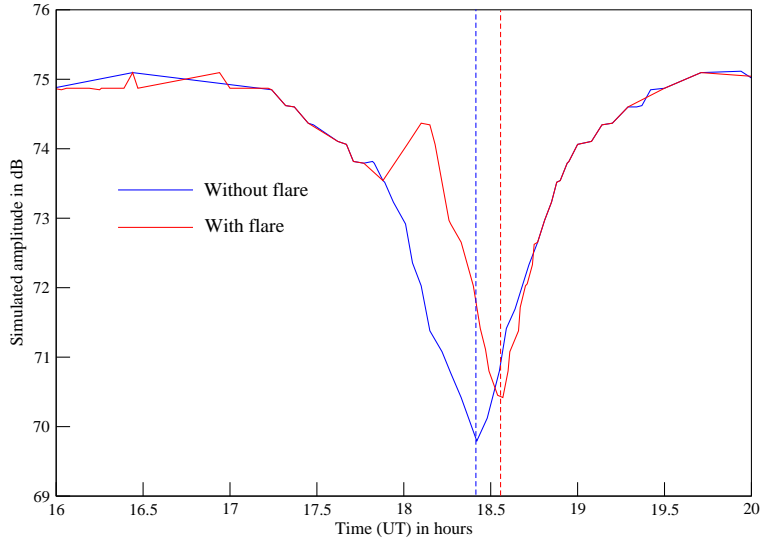}}
\caption{Simulated VLF signal amplitude as a function of time for NML-K5TD path. The blue curve represents the signal amplitude in the absence of 
the solar flare and the signal amplitude is modulated only due to the solar eclipse. The red curve represents the simulated amplitude where both the 
effects of solar eclipse and solar flare are included. The two dashed vertical lines represent the minimum value of signal amplitude.}
\end{figure}

\noindent{Next, the set of $h^\prime$ and $\beta$ were incorporated in the Wait's formula to calculate the electron density profile. Figure 7
shows the electron density profile at McBaine. Along x axis, we plot the time (UT) in hours and along y axis, we plot the ionospheric altitude in
km. The color-bar shows the electron density in cm$^{-3}$. From the Figure, we clearly see the electron density to decrease gradually as it 
approaches the time of maximum solar obscuration and thereafter it gradually increases ultimately returning to normal unperturbed value. The maximum
change in electron density is found to be $\sim 84\%$ at 80 km altitude.}\\

\begin{figure}[h]
\centering{
\includegraphics[width=10.0cm]{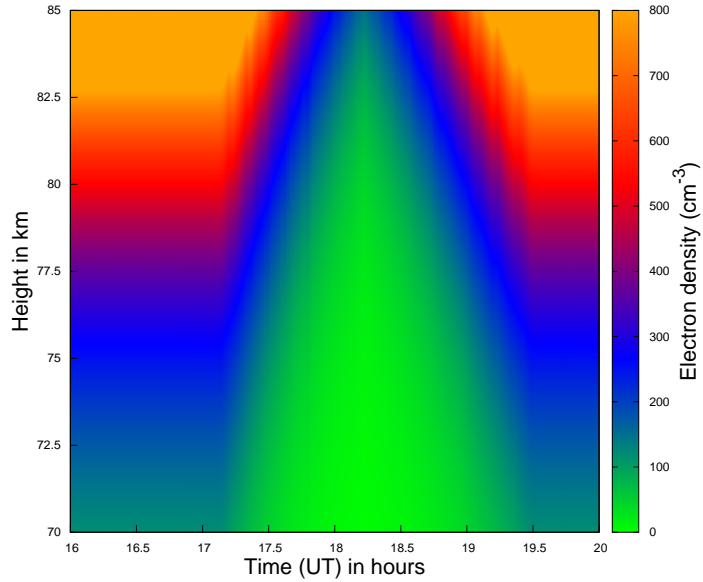}}
\caption{Altitude profile of electron number density (cm$^{-3}$) as obtained from LWPC and Wait's formula during the total solar eclipse at McBaine.} 
\end{figure}

\noindent{In Figure 8, we show similar results for Tulsa. The left column corresponds to the situation if there were no solar flare. In this
Figure, we find the variation of electron density to display a similar trend as for McBaine and the maximum deviation in electron density is found to
be $\sim 80\%$ at 80 km altitude. If we include the effect of the flare along with the eclipse as shown in the right panel figure, then the 
electron density profile exhibits a different nature of variation. In that case the electron density initially begins to decrease altogether but 
suddenly increases because of the inclusion of newly formed electrons due to the flare. As the effect dissipates, the electron density again 
begins to decrease and follow the same trend as it would in the absence of the flare.}
\begin{figure}[h]
\centering{
\includegraphics[width=6.0cm]{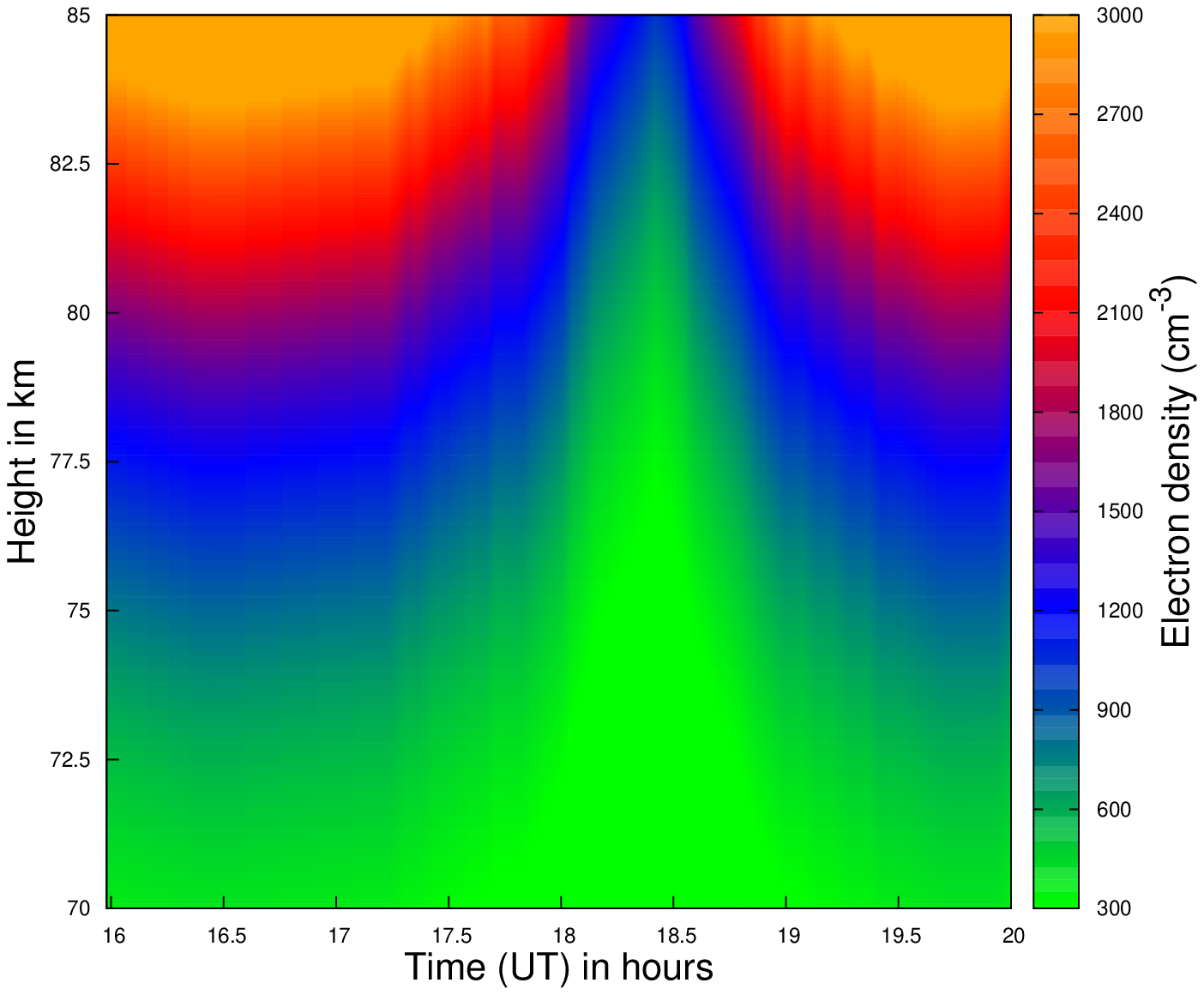}
\includegraphics[width=6.0cm]{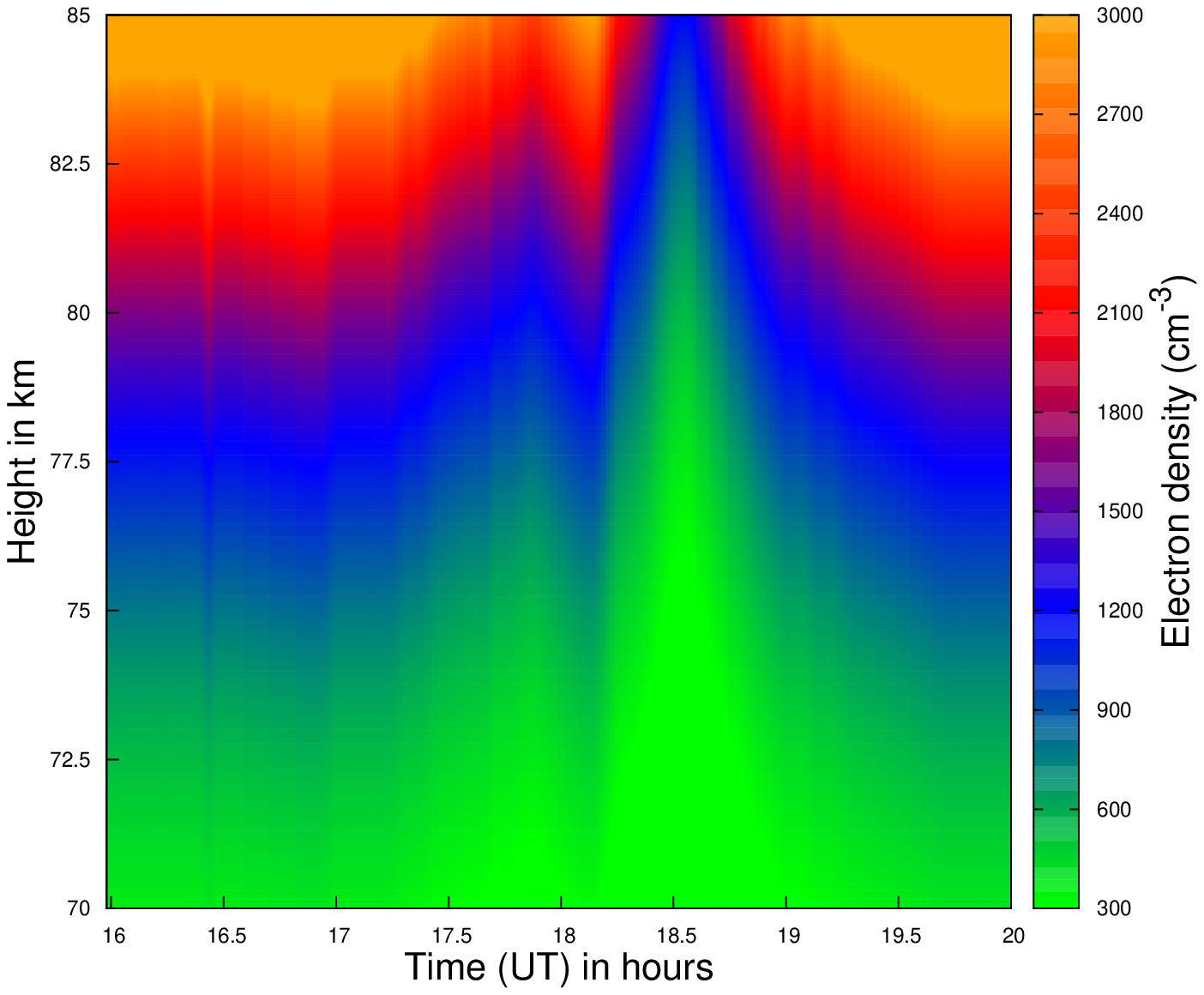}}
\caption{Altitude profile of electron number density (cm$^{-3}$) as obtained from LWPC and Wait$'$s formula at Tulsa due to solar eclipse alone 
(left column) and due to the combined effects of both solar eclipse and solar flare (right column).}
\end{figure}

\section{Conclusion and Discussion}

\noindent{In this paper, we wanted to study the compositional changes of the lower ionosphere during a total solar eclipse in August 21, 2017 which took 
place across USA. It was interesting that a flare also occurred during this period. The signal was recorded from numerous stations spread around the 
path of totality using two different monitors. In this study we chose data from two locations, namely, McBaine and Tulsa. McBaine was within 
the totality belt and the flare was always blocked by the lunar disk. Tulsa, on the other hand, was outside the totality belt but close to it with a 
maximum obscuration of 88.61\%. It experienced the obstructed flare. This provided us an excellent opportunity where we could study the ionospheric 
response to the eclipse, both including and excluding the additional complication due to the flare. To understand the physical mechanisms behind the 
ionospheric variabilities, we modeled it using several steps. First, we calculated an obscuration function due to the lunar disc that decides the 
variation of flux at a particular location on earth as a function of time. This gives us a temporal profile of the incoming ionizing radiation. 
Using a linear relation between this incident radiation and the ionospheric parameters, namely, the ionospheric reflection height $h^\prime$ and the 
steepness parameter $\beta$, we determined a set of their values that best replicated our observations. This set of parameters was then incorporated 
into the well known Long Wavelength Propagation Capability (LWPC) code to reconstruct the signal amplitude variation during the eclipse period. Later, 
the same parameters were used in the Wait's formula to obtain the electron density profile. For Tulsa, the whole process was iterated once more to 
include the additional effects due to the flare. The results obtained were quite satisfactory. The signal amplitudes at the two receiving locations 
displayed an overall opposite nature of deviation and the simulated results as depicted in Figs. 4 and 5 agreed to such observations to good extent. 
The times of occurrence of minima in the recorded signal amplitude also matched the simulation results. For Tulsa, the signal amplitude portrayed an 
unique feature where the signal began to decrease under the influence of the eclipse and then responded to the onset of the flare by beginning to 
regain its strength. Later after the effects of the flare dissipated, the signal variation behaved as though it was under the influence of the eclipse 
alone. The ionosphere is somewhat sluggish in its response to the ionization and the recombination processes and hence such observations can be 
attributed to this particular behavior of the ionosphere. Using this information into our numerical model, we are able to obtain results which roughly 
corroborate the observational findings (Fig. 6). From the modeling, we obtained yet another interesting result where we found that for the flare, the 
minimum in the signal amplitude shifted almost 12 minutes from where it would have been, without the presence of the flare. This result is a 
manifestation of the effects of the flare on the ionosphere and helps in validating our model more firmly. The electron densities as calculated were 
also found to vary accordingly and the maximum decrease in electron density at 80 km altitude is found to be $84\%$ for McBaine and 
$80\%$ for Tulsa.}\\

\noindent{It can be concluded that purely from modeling approach, and including all the realistic physical processes of the lower ionospheric region 
in our simulation, the observations could be reconstructed to a great extent. Obviously more improvement in the numerical model, particularly in 
modeling the effects of the solar corona could have delivered better agreement between observations and our model signal variation. However, the 
effects of corona on our lower ionosphere is poorly understood and any detailed modeling may not be able to unwarranted assumptions, which we wished 
to avoid here. Even then we believe that we have generally captured the salient features of the observations.}

\vskip 1.0cm
\textbf{Acknowledgement:}
The authors of this paper thank SuperSID network of Stanford University Solar Center for providing the VLF data during the solar eclipse by 
multi-station monitoring in North America. The authors thank Prof. Deborah Scherrer of Stanford Solar Centre which runs the SuperSID network. 
The authors also thank Dr. Sujay Pal for scientific discussions and Mr. Soujan Ghosh for handling the SuperSID data.




\end{document}